\documentstyle[12pt]{article}
 at 14.4truept at 12.0truept
\title{ 
The Non-Perturbative ${\cal N} = 2 $ SUSY Yang-Mills Theory from 
Semiclassical
Absorption of Supergravity by Wrapped D Branes} 
\author{
{\large Satchidananda  Naik}
\\
  Harish-chandra Research Institute \\
 Chhatnag Road, Jhusi  \\
Allahabad-211 019, INDIA\\}

\begin{document}
\maketitle

\hspace*{\fill}

\hspace*{\fill}
\newcommand{\bee}{\begin{equation}}
\newcommand{\nn}{\nonumber}
\newcommand{\ee}{\end{equation}}
\newcommand{\ba}{\begin{array}}
\newcommand{\ea}{\end{array}}
\newcommand{\bea}{\begin{eqnarray}}
\newcommand{\eea}{\end{eqnarray}}
\newcommand{\ki}{\chi}
\newcommand{\eps}{\epsilon}
\newcommand{\pa}{\partial}
\newcommand{\lb}{\lbrack}
\newcommand{\Se}{S_{\rm eff}}
\newcommand{\rb}{\rbrack}
\newcommand{\de}{\delta}
\newcommand{\th}{\theta}
\newcommand{\rh}{\rho}
\newcommand{\ka}{\kappa}
\newcommand{\al}{\alpha}
\newcommand{\bt}{\beta}
\newcommand{\si}{\sigma}
\newcommand{\bsi}{\Sigma}
\newcommand{\vp}{\varphi}
\newcommand{\gm}{\gamma}
\newcommand{\gb}{\Gamma}
\newcommand{\om}{\omega}
\newcommand{\et}{\eta}
\newcommand{\qab}{{{\sum}_{a\neq b}}{{q_a q_b}\over{R_{ab}}}}
\newcommand{\omb}{\Omega}
\newcommand{\pr}{\prime}
\newcommand{\ra}{\rightarrow}
\newcommand{\nb}{\nabla}
\newcommand{\MSb}{{\overline {\rm MS}}}
\newcommand{\lnh}{\ln(h^2/\Lambda^2)}
\newcommand{\cz}{{\cal Z}}
\newcommand{\h}{{1\over2}}
\newcommand{\Lm}{\Lambda}
\newcommand{\inft}{\infty}
\newcommand{\bpa}{\bar {\partial}}  
\newcommand{\hth}{\hat {\theta}}
\newcommand{\hp}{\hat p}
\newcommand{\hb}{\hat b}
\newcommand{\hc}{\hat c}
\newcommand{\hbt}{\hat {\beta}}
\newcommand{\hgm}{\hat {\gamma}}
\newcommand{\hvp}{\hat {\varphi}}
\newcommand{\hlm}{\hat {\lambda}}    
\newcommand{\hv}{\hat v} 
\newcommand{\hq}{\hat q} 
\newcommand{\Lra}{\Longleftrightarrow}
\newcommand{\abschnitt}[1]{\par \noindent {\large {\bf {#1}}} \par}
\newcommand{\subabschnitt}[1]{\par \noindent
                                          {\normalsize {\it {#1}}} \par}
%-----------------------------------------------------------------------
% The definition below makes spaces e.g \skipp{3} makes 3 spaces
\newcommand{\skipp}[1]{\mbox{\hspace{#1 ex}}}
 
%
%
% various slashed symbols
%
%
%\newcommand\slash#1{\rlap{$#1$}/} % slashes a character
\newcommand\dsl{\,\raise.15ex\hbox{/}\mkern-13.5mu D}
    % this one can be subscripted
\newcommand\delsl{\raise.15ex\hbox{/}\kern-.57em\partial}
\newcommand\Ksl{\hbox{/\kern-.6000em\rm K}}
\newcommand\Asl{\hbox{/\kern-.6500em \rm A}}
\newcommand\Dsl{\hbox{/\kern-.6000em\rm D}} %roman D
\newcommand\Qsl{\hbox{/\kern-.6000em\rm Q}}
\newcommand\gradsl{\hbox{/\kern-.6500em$\nabla$}}
%--------
 %---------------------------------------------------------------
 %\vskip5.0cm
\newpage
\begin{abstract} \normalsize
The imaginary part of the two point functions of the superconformal 
anomalous currents are extracted from the cross-sections of semiclassical
 absorption of dilaton, RR-2 form and gravitino by the wrapped D5 branes.
From the central terms of the two point functions anomalous Ward identity
is established which relates the exact pre-potential of the ${\cal N}=2$ SUSY 
Yang-Mills theory with the vacuum expectation value of the anomaly 
multiplet. From the Ward identity, WDVV 
(Witten-Dijkgraaf-Verlinde-Verlinde) equation can be derived which is 
solved  for the exact pre-potential.
\end{abstract}

\vskip10.0cm
\newpage
\pagestyle{plain}
\subabschnitt{1. Introduction:}
Recently the gauge theory/gravity duality , which is commonly
known as Ads/CFT duality is extended to non-conformal pure 
 ${\cal N}=1$ or ${\cal N}=2$  supersymmetric Yang -Mills (SYM)
theories \cite{Mald}. So far the perturbative behaviour of the ${\cal 
N}=2$ SYM
is produced by this duality. The conventional folklore is that the 
instatons which are responsible for the non-perturbative part of
the pre-potential are suppressed in the large N limit (since the
gauge gravity duality is valid only in the large N limit). Since  the 
wrapped D5 branes are far from being  conformal, the usual
holographic description of bulk-boundary connection is quite  subtle to
extract all the properties of SUSY Yang-Mills theories. This fact led us 
to wander other ways of getting complete picture of the non-perturbative
solution of the SUSY Yang-Mills from gravity. It is a fact in quantum 
field theory that the absorption/emission cross section of particles is 
always given by the discontinuties of  the two point functions of the 
currents to which the field couples. To be precise, if the interaction 
Lagrangian is $S_{\rm int} = \int d^4 x \,{\varphi}(x)\,{\cal J}(x)$ 
then the
cross section is 
\bee
 \sigma = {1 \over 2 i \omega} Disc~ \Pi (p)
    \bigg|_{p^0 = \omega \atop \vec{p} = 0} \ ,
\ee
where
\bee
\Pi (p) = \int d^4 x \, e^{i p \cdot x}
    \langle {\cal J}(x) {\cal J}(0) \rangle .
\ee
The connection between the central charge of four dimensional energy 
momentum tensor and the absorption of gravitons/dilatons by the world 
volume theory of D branes is first explored by Gubser and 
Klebanov \cite{GK}. The trace of the energy-momentum tensor, the $\gm$
 trace of the super current and the
 divergence of $U(1)$ axial current due to the
$R$-symmetry form different components of the superconformal anomalous 
current $J_{\alpha {\dot{\alpha}}}$ and its super divergence 
${\bar D}^{{\dot{\alpha}}}J_{\alpha {\dot{\alpha}}}$ is classically zero
however possesses anomaly. More explicitly
\bea
\langle {\sqrt g}\,\, {\theta}^\mu_\mu \rangle 
&&   = \h
\frac{\bt(g)}{g^3}\left(F^a_{\mu\nu}\right)^2 ~ +
\frac{c(g^2)}{16{\pi}^2} \left(W_{\mu\nu\rho\si}\right)^2 ~ -
\frac{a(g^2)}{{16{\pi}^2}}\left({\tilde R}_{\mu\nu\rho\si}\right)^2\nn\\
\langle {\pa }^\mu {{\sqrt g} R}_\mu \rangle
&& = - \frac{\bt(g)}{3g^3}\left(F^a_{\mu\nu}{\tilde F}^{a 
\, {\mu\nu}}\right) + \frac{c(g^2) - a(g^2)}{{24{\pi}^2}}\left({\tilde 
R}R\right)\nn\\
\langle {\sqrt g} {\gamma}^\mu {\cal S}_\mu \rangle &&=
\h\frac{\bt(g)}{g^3}\left[ \left({\si}_{\mu\nu}{\lambda}^a 
F^{a\mu\nu}\right) + \h {R}^{\mu\nu\rho\si}{\si}_{\rho\si}\left(D_{\mu}
{\psi}_{\nu} - D_{\nu}{\psi}_{\mu}\right) - 
\left({\epsilon}^{\mu\nu\rho\si}
{\gamma}_5 {\psi}_{\mu}A^a_{\nu}F^a_{\rho\si}\right)\right].\nn\\
\eea
where $\bt(g)$ is the beta function of SYM, $a(g)$ and $c(g)$ are 
central
functions near the criticallity, $W_{\mu\nu\rho\si}$ is the Weyl tensor
and ${\tilde R}_{\mu\nu\rho\si}$ is the dual of the curvature tensor
and $\psi_{\mu}$ is the gravitino field.
We first derive the world volume action of the wrapped D5 brane where
these  currents are minimally  coupled to the ten dimensional 
supergravity fields. Then we calculate the absorption cross-section of 
these fields by the brane which give us the central functions of these
correlators.
The organization of this paper is as follows. First we describe briefly 
the preliminaries of the wrapped D5 brane which gives the ${\cal N} = 2$
SUSY Yang-Mills theory \cite{gaunt,div}. Then we show the absorption 
of 
dilaton, fluctuating part of RR-2 forms and the gravitino by the world 
volume of wrapped D5 branes and evaluate the central functions. 
Consequently we derive the equation relating the pre-potential with the
vacuum expectation of the anomaly multiplet and then conclude.

\subabschnitt{2. The Preliminaries:}
We start with  type IIB little string theory e la' a collection of a 
large number of $NS5$ brane in the vanishing string coupling limit
which gives rise to $D =6$ SYM \cite{Seib}. Then we dimensionally reduce 
two of its spatial world volume in such a way that we retain ${\cal N}=2$ 
SYM  in the low energy limit \cite{cv}. The $NS5$ brane has $SO(4)$ 
R-symmetry
as the normal bundle. When one identifies the $U(1)$ subgroup of the
$SO(4)$ R-symmetry with the $U(1)$ spin connection of the two cycle which 
is compactified, one gets a covariant constant spinor and SUSY is 
retained, which is commonly known as twisting \cite{Mald}.
This is called wrapping of $NS5$ brane on a supersymmetric two cycles.
If the compact space is a two-shere, then there will be no extra hyper- 
multiplet and in the low energy limit i.e. in the scale much lower than
the radius of the sphere we will get pure  ${\cal N}=2$  SYM.  
 Thus it 
amounts to consider a gauged $D = 7$ supergravity solution and then lift 
it to get the solutions in ten dimensions. In recent past many relevant 
classical solutions have been done which  we don't want to repeat 
instead
 we use here the results of
\cite{gaunt,div} classical solutions of $D = 7$ gauged supergravity
which is amenable to ten dimensional string theory.
\bea
ds_{10}^2 &=& {\rm e}^{\Phi}\Bigg[ d x_{1,3}^2 +
\frac{z}{\lambda^2} \left( {d{\theta}}^2 +
\sin^2 {\theta} \,{d {\varphi}}^2 \right)
+\frac{1}{\lambda^2} \,{\rm e}^{2x} \,dz^2 \nn \\
&&~
+ \frac{1}{\lambda^2} \left(d {\theta}_1^2 +
\frac{{\rm e}^{-x}}{f(x)} \cos^2 {\theta}_1 \left(d
  \theta_{2} + \cos {\theta} \,d {\varphi}
\right)^2
+ \frac{{\rm e}^{x}}{f(x)} \sin^2 {\theta }_1\,d \theta_{3}^{2}
 \right)\Bigg]~~,
\eea
where the dilaton is 
\bee
{\rm e}^{2\Phi} = {\rm e}^{2 z} \left[1 - \sin^2 
{\theta}_1 \frac{1 + c \,{\rm e}^{-2 z}}{2  z} \right]~~
\ee
and
\bee
f(x) = {\rm e}^{x} \cos^2 {\theta }_1 + {\rm e}^{-x} \sin^2 {\theta}_1,
\ee
also
\bee
 {\rm e}^{-2x} = 1 - \frac{1+c \,{\rm e}^{-2 z}}{2  z}
\ee
where $\lambda$ is the  gauge  coupling constant of seven dimensional 
gauged supergravity and $c$ is a parameter as the integration constant
of the classical solution. For $c\geq -1$ the range of the radial 
variable is $z_0\leq z \leq \infty$ where $z_0$ is the solution
for $ {\rm e}^{-2x(z_0)} =0$. Here $\theta$ and $\varphi$ are the
angles of compact two-sphere with radius of compactifaction as 
$\frac{z}{\lambda^2}$
and $\theta_1$, $\theta_2$ and $\theta_3$
are angles of transverse three-sphere.
The conservation of the RR-charge on
the transverse sphere $S_3$ fixes $\frac{1}{\lambda^2}= N\,g_s\,\alpha'$
for  large  $N$ number of $D5$ branes with string coupling $g_s$.
The R-R- 2 Form is given by 
\bee
C^{(2)} = \frac{1}{\lambda^2}\, {\theta}_3\,d 
\left[\frac{\sin^2{\theta}_1}
{f(x)\,{\rm e}^{x}} \,( d {\theta}_2 + \cos
{\theta}\, d {\varphi}) \right].
\ee
The $D5$ brane action is given by
\bee
S = - \tau_5 \int d^6 \xi ~{\rm e}^{- \Phi}
\sqrt{- \det \left( G+ 2 \pi \alpha' F \right)} + \tau_5
\int\left(\sum_n C^{(n)}\wedge {\rm e}^{2 \pi \alpha' F}
\right)_{6-{\rm form}}
\ee
where F is the world volume gauge field and $\tau_5$ is the brane 
tension.
. The BPS condition is fixed from 
the condition of the vanishing of the potential between two branes which 
gives $\theta_1$ to be $\frac{\pi}{2}$. This condition makes the 
transverse boundary of the D brane to be a two dimensional space
consisting $z$ and $\theta_3$ which will eventually the moduli space
of  ${\cal N}=2$  SYM.
\subabschnitt{3. The Absorption of Dilaton:}\noindent
To evaluate the central terms for the correlation function of stress 
tensor we need to calculate the absorption of the fluctuating dilaton 
coupled to the world volume of the wrapped D5 brane. If we denote the
fluctuation of dilaton by $\eta$ then the world volume action is given 
by
\bee
S_{\rm int} =  {\tau}_5 \int d {\Omega}_2\, \int d^4 x \,
 {\rm e}^{-\Phi}\, \sqrt{-\det G}~ \,\eta (x,\xi) \,
 \left\{ \frac{1}{4}
 F_{\alpha \beta}^a \,F^{a \alpha \beta} +\,\h  {\overline\lambda}^a 
{\gm}^{\alpha}D_{\alpha}{\lambda}^a +\, 
 \frac{1}{2} D_{\alpha}
 {\overline\Psi}^a\, D^{\alpha} \Psi^a \right\}.
\ee

Here $F^{a \alpha \beta}$ is field strength of the SUSY Yang-Mills,
${\lambda}^a$  are the fermions and $\Psi^a$ are the scalars of 
${\cal N}=2$ multiplet in the adjoint representation of the $U(N)$
gauge group. Here $\Psi^a =\, T^a\, {\rm e}^{(z + i {\theta}_3)}$ is the 
complex scalar which belongs 
to the transverse space of the $D5$ brane which 
eventually forms the moduli space of the ${\cal N}=2$ SUSY Yang-Mills
vacuua and $T^a$ is the generator of the $U(N)$
. It has been shown by Klebanov \cite{kl} and Gubser et 
al.\cite{gka}
that the string theoretic abosorption cross-section and the classical
absorption cross-section given by the classical equation of motion of
dilaton  by the $D3$ brane  are in exact agreement. The classical 
absorption probability is given as the ratio of the flux at infinity to
the flux where the brane is sitting \cite{ma}. The classical equation 
of motion by 
the dilaton is
\bee
 {\pa}_a{\sqrt G}G^{ab}{\pa}_b {\vp} = 0.
\ee
where $G^{ab}$ is given by the metric given in eq.(4). If we define the
transverse coordinate ${\rm e}^z = \rho$ then for large $\rh$, 
$G^{\rh\rh}= \,{\rh}$ and ${\sqrt G} = {\rh}^2\, R^4$ where $R^2$ is the 
square of the
compactified volume which is identified with $\frac{1}{\lambda^2}$. The
 equation of motion for the  dilaton is
\bee
{\pa}^2_{\rh} {\vp} +\, \frac{3}{\rh}{\pa}_{\rh}{\vp} 
+\,\frac{R^2\om^2}{{\rh}^2} = 0.
\ee
We take the ratio of the flux ${\vp}^*\, G^{\rh \rh}\, 
{\pa}_{\rh}\,{\vp}$ for
$\rh ~ > >~ \om R$ to that for  $\rh ~ \leq  \om R$. This is
the probability of absorption which multiplied with the proper
phase space gives the cross section. This is 
 also 
same as quantum cross section of the world volume theory by the normalized
dilaton fluctuation $\eta$ c.f. eq.(10).
The cross -section is
\bee
d\si = \frac{1}{2\om}\,
\frac{d^5p_1}{(2\pi)^52E_1} \,\frac{d^5p_2}{(2\pi)^52E_2}\,(2\pi)^6\,
 {\delta}^6\left(p_1 + p_2 - q\right)\,                                     
{\overline{|{\cal M}|}}^2_{fi}
\ee
where ${\cal M}_{fi}$ is the matrix element for  the dilaton with 
momenum q  going
to two gluons, two gluinos and two  scalars with momenta $p_1$ and $p_2$.
We take the  vector component of the momenta to be zero and the zeroeth 
component of the dilaton to be $\om$. The momentum conservation gives
$E_1 + E_2 = \om$. After the integration of the phase space the 
cross-section is found to be
\bee
\si = {\frac{V_2}{16}}\frac{{\kappa}^2_{10}}{32{\alpha}^{\pr} \pi}{\om}^3
\ee
where $V_2$ is the volume of the compact two dimensional space
which arises due to the delta function in these directions. Here
$V_2 = 4\pi R^2 \ln\rh$. We identify $2\pi g_s\, =\, g^2_0$
the Yang-Mills coupling.
Then 
\bee
\si = \frac{{\kappa}^2_{10}}{32 \pi}{\om}^3
\left(\frac{g^2_0 N\, \ln\, \rh}{8{\pi}^2}\right)
\ee
This gives the imaginary part of the  polarization $\Pi(p)$.(c.f. eq.(2)).
It has been shown by Anselmi et al.\cite{ans} how to get all the central
functions of eq(3). from the two point correlation functions. Any two 
point 
function of currents
\bea
 {\cal G}(x) &=& \,\langle {\cal J}(x) {\cal J}(0) \rangle\nn \\ 
             &=& \, \frac{b(\ln(x\mu))}{x^{2d_0}}
\eea
where $d_0$ is the engineering dimension of the current ${\cal J}$
and $\mu$ is a  scale parameter. The 
two point function of the local operators aught to satisfy the 
Callan-Symanzik equation,
\bee
\left( x {\pa\over {{\pa}x}}\,+ 2d_0 + 2 {\gm}_0 +\, \bt (g){\pa\over 
{{\pa} g}}\right){\cal G}(x) = 0.
\ee
where  $\gm$ is the anomalous dimension of the operator and $\bt(g)$ is 
the beta function of the renormalization group.
From eq.(15 ), eq.(16) and eq.(17) we get 
\bee
\bt(g_{\rm YM}) =  -\frac{N}{8\pi^2}\,g_{\rm YM}^3~.
\ee
\subabschnitt{4. The $U(1)_R$ Anomaly:}\noindent
The gaugino and the scalars of  ${\cal N}=2$ SUSY Yang-Mills theory 
possess $U(1)$ R-Symmetry. For the wrapped branes it is realized as the 
rotations in the transverse space mainly by the angle ${\theta}_3$. 
However this is periodic and a change of it needs to be discrete
like $\frac{\pi}{N}\,k$ since R-R 2-Form flux should be quantized
on the vanishing  compact  two sphere. One realizes 
the chiral anomaly due to this fact in the gravity dual picture.
We want here to calculate explicitly the two point function of the chiral
R-current in the world volume theory of wrapped D5 branes.
The fermionic partner of the Chern-Simon's term of the second part of 
eq.(9) is
\bee
S_{\rm int} =  {\tau}_5 \int d^6{\xi}~\,{\om}_2\,\wedge 
{\epsilon}^{\mu\nu\lambda\si}\,{\Pi}^a_{\mu}\,{\Pi}^b_{\nu}
\,{\Pi}^c_{\lambda}\,
\,{\overline\lambda}~{\gb}_{abc}~{\lambda}{\pa}_{\si} c(x)
\ee
where we have used RR-2 form $C^2 = {\om}_2~ c(x)$ and $c(x)$ is the 
fluctuation of $C^2$. Here ${\Pi}^a_{\mu}(x)~=~ {\pa}_{\mu}X^a - 
i{\overline\th}{\gb}^a{\pa}_{\mu}{\th}$ is like vielbein in the 
Green-Schwarz variable. Here $\{a ,b = 0,..,10\}$ and $\{ \mu, \nu = 
0,..,3\}$
. Integrating over ${\om}_2$ and using the 
properties of the $\gb$ matrices we get
\bee
S_{\rm int} ~=~  N \int d^4x\, {\overline\lambda} i~ {\gm}_5~ 
{\gm}_{\mu}~{\lambda} 
       \,  {\pa}^{\mu}c(x)\,.
\ee
Here $c(x)$ is a fluctuating field which is classically ${\th}_3$.
 We want 
to calculate the classical absorption croos-section of $c(x)$ by the 
wrapped D5 brane. The classical action for this field is extracted
from the low energy limit of the type IIB super string action.
\bea
S &=& {\frac{1}{2 k^2_{10}}}  \int d^{10} x \bigg({\sqrt{-g_{10}}}
 \big[ R -\, \h {(\pa \phi)}^2
- { \textstyle{1\over 12}} e^{-\phi}  { (\pa B_2)}^2
 -  \,  { \textstyle{1\over 2}}  e^{2 \phi} (\pa C)^2\nn\\
 & - &{ \textstyle{1\over 12}} e^{ \phi}  {(\pa C_2  - C  \pa B_2)}^2
- { \textstyle{1 \over{4\cdot 5!}}}  F^2_5\,\big]\bigg)\\
\eea
Here $B_n$ and $C_n$ are n-Form potential due to NS-NS and R-R sectors 
respectively. We set $C_0 = C $ and  $F_5$ to zero and also $C_2 = - 
B_2$.
Substituting $F^{\theta\,\vp} = sin {\theta}$ from eq.(8) in
equ.(12) and integrating  over $\theta$ 
and $\vp$ coordinates 
we get the action for $c(x)$ which effectively moves in six dimension as
\bee
 S = { \textstyle{1\over 6}}{\frac{1}{R^2\,\ln {\rh}_0  k^2_{10}}}
\int d^{6} x \bigg({\sqrt{-g_{6}}}{\pa}_a  c(x)\,g^{ab} {\pa}_b 
c(x)\bigg).
\ee
The classical absorption probability  can be calculated as the ratio
of the  in coming flux for $\rh $ at infinity to the outgoing flux
from $\rh {\rm < }R\om $. Also from  the world volume action the 
absorption cross-section is
\bee
\si ~= ~ \int \frac{d^3p_1}{(2\pi)^32E_1} 
\,\frac{d^3p_2}{(2\pi)^32E_2}\,(2\pi)^4\,
 {\delta}^4\left(p_1 + p_2 - q\right)\,
{\overline{|{\cal M}|}}^2_{fi}
 \ee
where ${\cal M}_{\rm fi}$ is given as 
\bee
{\cal M}_{\rm fi}\, =\, -{\frac{1}{2}} \sqrt{2 \,{\tilde \kappa}_{10}^2}
    q^{\mu} \bar{v}(-p_1) (i \gamma_5~ {\gamma}_{\mu}~u(p_2)
\ee

where ${\tilde \kappa}_{10}$ is the normalization for $c(x)$ which is 
from eq.(23),
\bee
{\tilde \kappa}^2_{10} = R^2\,\ln {\rh}_0  k^2_{10}.
\ee
This gives the cross-section
\bee
\si = \frac{{\kappa}^2_{10}}{32 \pi} {N}^2{\om}^3
\left(\frac{g^2_0 N\, \ln\, \rh}{24{\pi}^2}\right)
\ee
Here again the Callan-Symanzik equation gives the correct $\bt$ 
function.
\subabschnitt{5. The $\gm$ trace of the Super Current and Anomaly:}
The world volume action of the minimally coupled gravitino 
${\psi}_{\mu}$
to the wrapped $D5$ brane is
\bee
S_{\rm int} =  {\tau}_5 \int d {\Omega}_2\, \int d^4 x \,
 {\rm e}^{-\Phi}\, \sqrt{-\det G} \,\left\{ \frac{1}{4}
 {\psi}_{\mu}\,{\gamma}^{\mu}\,{\si}_{\al \bt}
{\lambda}^a~ F^{a \alpha \bt}\,\right\}
\ee
The gravitino part of the  low energy effective action for the type 
IIB string theory is 
\bee
S_{\rm gravitino} = {\frac{1}{2 k^2_{10}}}  \int d^{10} x 
\bigg({\sqrt{-g_{10}}} \h {\overline\psi}_{\mu}\,i 
 {\gb}^{\mu\nu\rh}D_{\nu} {\psi}_{\rh}\bigg).
\ee
Due to twisting the covariant derivative $D_{\nu}$  is like  ordinary
one here.
For the classical equation of motion of the gravitino we choose
  a gauge ${\psi}_{\mu} = {\gm}_{\mu}\chi$ which simplifies the
equation of motion for the gravitino as
\bee
     {\rh}{\gm}_5\, {\pa}_{\rh} \chi +\, R {\gm}_0{\pa}_0 \chi = 0.
\ee
The probability of absorption is the ratio of the flux 
${\overline\chi}{\gm}_5 {\chi}$ at $\rh{\rm >>} R\om$ to 
$\rh{\rm <} R\om$. The absorption cross-section from the world volume
theory of minimally coupled gravitino gives the same result.
The phase space volume which comprise the compact two dimensional
sphere gives the $\ln {\rh}_0$ as the prefactor which eventually
provides the running coupling and the $\bt$ function. We get the same 
beta function as the other cases.
%%%%%%%%%%%%%%%%%%%%%%%%%%%%%%%%%%%%%%%%%%%%%%%%%%%%%%%%%%%%%%%%%%%%%%%%%
\subabschnitt{6. The exact Pre-Potential:}
The  low energy effective action for ${\cal N}$ = 2 SUSY Yang-Mills 
theory is 
\bee
{\gb}\big( \Phi\big) \,=\, \int d^4x d^4{\th}{\cal F}
\big( \Phi\big)
\ee
where ${\cal F}$ is the pre-potential and ${\Phi}$ is the chiral
superfield of the ${\cal N = 2}$ SUSY Yang-Mills theory. To realize 
the 
superconformal invariance a supergravity prepotential $H^{\alpha 
{\dot{\alpha}}}$ is coupled to the superconformal current $J_{\alpha 
{\dot{\alpha}}}$. The invariance of the effective potential under
superconformal transformation up to first order gives the anomalous 
superconformal Ward identity\cite{howe}
\bee
\int d^4x\, d^4{\th}\,{\frac{\delta{\gb}\big({\Phi}\big)}
{\delta{\Phi}}}\,{\rm <}{\delta{\Phi}}{\rm >}
\,=\,\int d^4x\, d^4{\th}\delta H^{\alpha{\dot{\alpha}}}{\rm<}
J_{\alpha{\dot{\alpha}}}{\rm >}.
\ee
The righthand side of this equation gives 
$D^{\dot{\alpha}}J_{\alpha{\dot{\alpha}}}$ which is the divergence of 
the anomalous superconformal current and its vacuum expectation gives 
the anomaly,
\bee
2 {\cal F} - {\cal F(A) }'{\cal A} = \frac{N}{8\pi^2}\langle
tr{\upsilon}^2\rangle,
\ee
where ${\cal A}$ is the vacuum expectation value of ${\Phi}$
and  $tr{\upsilon}^2$ is the anomaly multiplet
for example $trF^2$ will correspond to ${\theta}^{\mu}_{\mu}$ etc.
Indeed the second order variation of the infinitesimal
supeconformal parameter ${\epsilon}^{\alpha}$ will relate  the
two point function of the derivative of the  anomalous currents
 which has been obtained in the last sections
to the two point function of the ${\Phi}$
with the derivatives of the pre-potential as the prefactor.
This identity is difficult to realize in the coulomb phase when
${\Phi}$ has a vacuum expectation value. So we resort here to
the Ward identity for our purpose. In our case ${\Phi}(x) = 
\Psi (x)$ . We assume here that all the D branes are distributed
on a circle with $N$ points in the ${\th}_3$ directions. So 
$\langle{\Psi}_i \rangle\, = \,{\rh}{\rm e}^{i{\th}^i_3}$ which
is denoted as ${\cal A}_i$. In this phase the Ward identity
\bee
 2 {\cal F} - {\sum}_i\frac{\pa {\cal F}}{\pa {\cal A}_i}{\cal A}_i
 \, =\,\frac{N}{8\pi^2}\langle tr{\upsilon}^2\rangle,
\ee
which is same as eq.(3) in the flat space. This equation can be 
rewritten as WDVV equation \cite{WDVV}. This is also same as the 
renormalization group equation of D' Hoker et al.\cite{dh} where the 
right 
hand side is given as ${\mu}\frac{\pa}{\pa \mu}$ of ${\cal F}$.
To solve for ${\cal F}$ the usual ansatz is to assume ${\cal F}$ as a
genus $N-1$ hyper-elliptic curve, 
\bee
y^2\,=\, P(x)^2 - {\Lambda}^N
\ee
with canonical homology basis $\{a_i, b_i\}$ and the differential form
${\cal S}_{\rm SW}~=~\frac{x~dP}{2 i~\pi~ y}$. 
The large N solution of this 
equation was first solved by Douglas and Shenker \cite{doug} and 
recently many subtlities are addressed  by Ferrari \cite{ferrari}. 
The non-perturbative nature of the theory will be studied from the 
distribution of the roots of eq.(35). The singularity occurs when any 
two of the roots coincide, where a dyon becomes massless. Our moduli 
space coordinates are ${\cal A}_i \,=\,{\rh}{\rm e}^{i{\th}^i_3}$ and
${\th}^k_3 = 2\pi\frac{k}{N}$. The differnce of ${\cal A}_i$ is the
W-meson mass and in their vanishing limit mass less particle emerges
signalling phase transition. The integration over the homology cycle
for example $a$   gives
\bee
i\pi\, {\frac{a}{N}}\, = \,-{\rh -1\over \rh}~ +~ \ln \rh
 +~ {1\over N}\, \left(-\ln \rh + {\rh-1\over \rh}\, \ln 2 + 
{(\rh-1)\ln(\rh -1)\over \rh}\right),
\ee
which shows clearly the singularity for $\rh \rightarrow 1$.
Indeed in  our supergravity picture the radius 
of the wrapped two sphere  $z ~= ~\ln \rh$   vainishes and 
signals "enhancons" \cite{Polch}. 
\subabschnitt{7. The Conclusion:}
Here we tried to analyze the non-perturbative 
effects of ${\cal N}$ = 2
SUSY Yang-Mills theory  starting from the absorption of
minimally coupled Supergravity fields by the wrapped $D5$ branes.
The absorption cross-section gives the two point correlators of the
currents. This is very crucial for our entire analysis which provides
the central functions. This encodes the entire evolution of the 
coupling dictated by the renormalization group. Once we get the beta 
function correctly we can use it to establish the anomalous Ward
identity. It is the cosistency of the renormalization group which
gives one point function from two point function. This analysis
  perhaps  can be 
obtained by renormalization group flow ${\cal N}$ = 4 to 
${\cal N}$ = 2  \cite{pw} which we found to be quite non-trivial
to achieve this goal.
%%%%%%%%%%%%%%%%%%%%%%%%%%%%%%%%%%%%%%%%%%%%%%%%%%%%%%%%%%%%%%%%%%%%%%

 \end{document}